\documentclass[aps,prl, superscriptaddress, twocolumn]{revtex4}  
\usepackage{amssymb}
\usepackage{amsmath}
\usepackage{bbm}
\usepackage{graphicx}
\usepackage{color}

\begin{document}

\title{Light-induced hidden odd-frequency order in a model for A$_3$C$_{60}$}
\author{Philipp Werner}
\affiliation{Department of Physics, University of Fribourg, 1700 Fribourg, Switzerland}
\author{Yuta Murakami}
\affiliation{Department of Physics, Tokyo Institute of Technology, Meguro, Tokyo 152-8551, Japan}
\date{\today}

\hyphenation{}

\begin{abstract} 
Laser driving in systems with competing or coupled electronic orders can lead to the enhancement of orders, or even to the appearance of hidden phases without an equilibrium analogue. Here we consider a model for A$_3$C$_{60}$ which exhibits a unique interplay between conventional and odd-frequency (or composite) orders. In particular, we show that photo-doping of the antiferromagnetic Mott insulating phase, as realized in Cs$_3$C$_{60}$, results in a paramagnetic gapped state with broken orbital symmetry. This hidden phase, which does not exist under equilibrium conditions, can be interpreted as an odd-frequency orbital-ordered state, and is conceptually related to the equilibrium Jahn-Teller metal in more weakly correlated compounds. Our study demonstrates the appearance of pure odd-frequency order via the nonthermal melting of magnetic order, and provides an interesting example of nonequilibrium control of electronic orders in a multi-orbital system. 
\end{abstract}


\maketitle

{\it Introduction.}
Intriguing prospects of nonequilibrium condensed matter physics are light-control of correlated systems and the exploration of hidden phases that do not exist in equilibrium~\cite{Giannetti2016,delaTorre2021}.
Prominent examples include photo-induced superconducting-like states \cite{Fausti2011,Kaiser2014,Mitrano2016,Buzzi2020} and metastable 
structures with associated metal-insulator transitions~\cite{Ichikawa2011,Stojchevska2014,Cho2016}. 
Strongly correlated materials with several active degrees of freedom (spin, charge, orbital, lattice), which typically host competing or coexisting orders~\cite{Dagotto2005}, provide an ideal playground for these investigations.  
For example, superconductivity has been induced via the laser-driven suppression of a competing stripe-order \cite{Fausti2011}, while polaronic Mott insulators can be switched to long-lived nonthermal conducting states \cite{Stojchevska2014,Cho2016}. Theoretical investigations have revealed hidden magnetic, orbital, and spin-orbital orders \cite{Li2018,Werner2020} in photo-doped multiorbital systems.   
Still, there are many open questions concerning nonequilibrium control of competing orders, the nature of hidden phases, and the mechanisms which stabilize them. 

\begin{figure}[t]
\begin{center}
\includegraphics[angle=0, width=0.49\columnwidth]{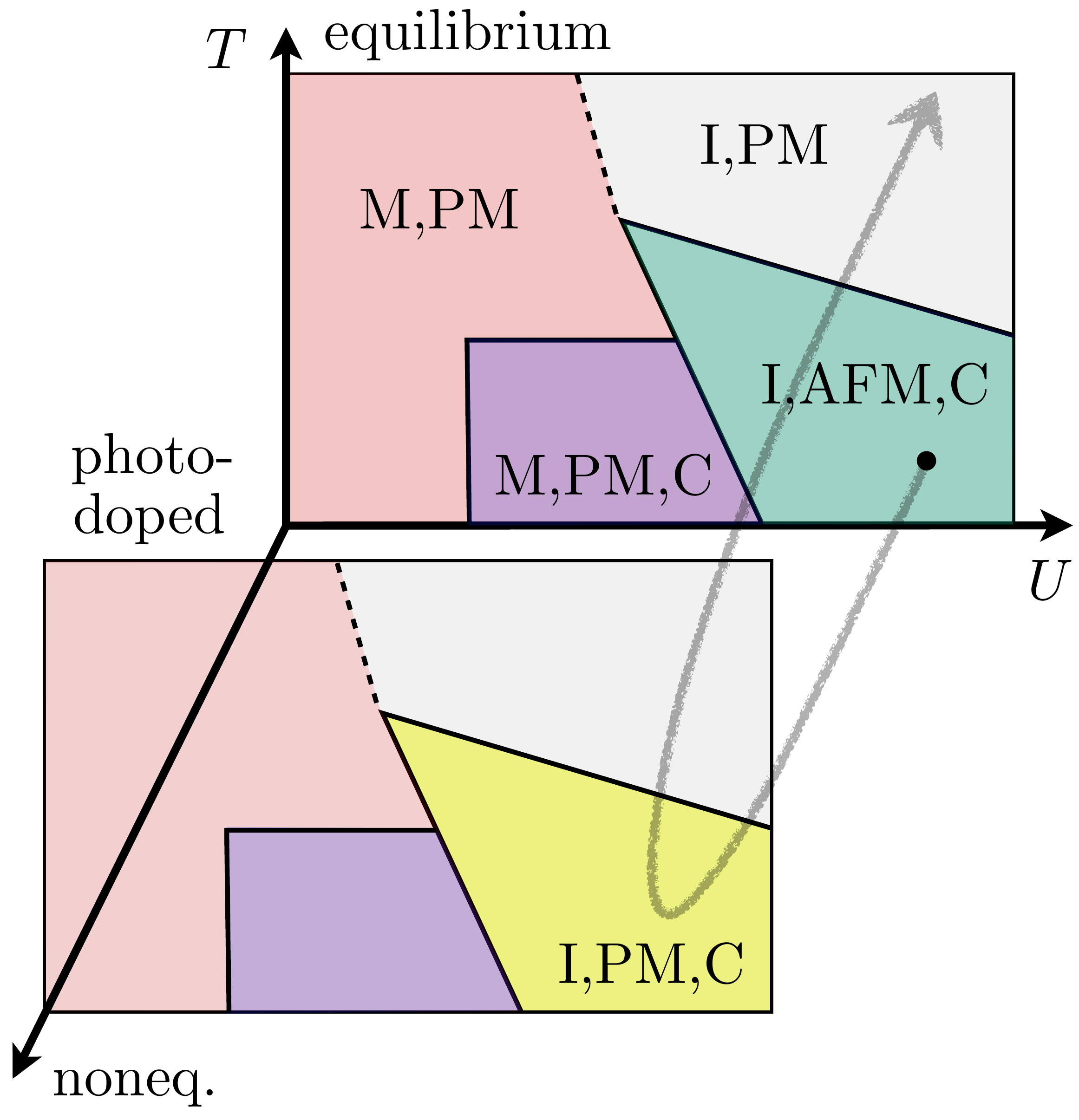} \hfill
\includegraphics[angle=0, width=0.49\columnwidth]{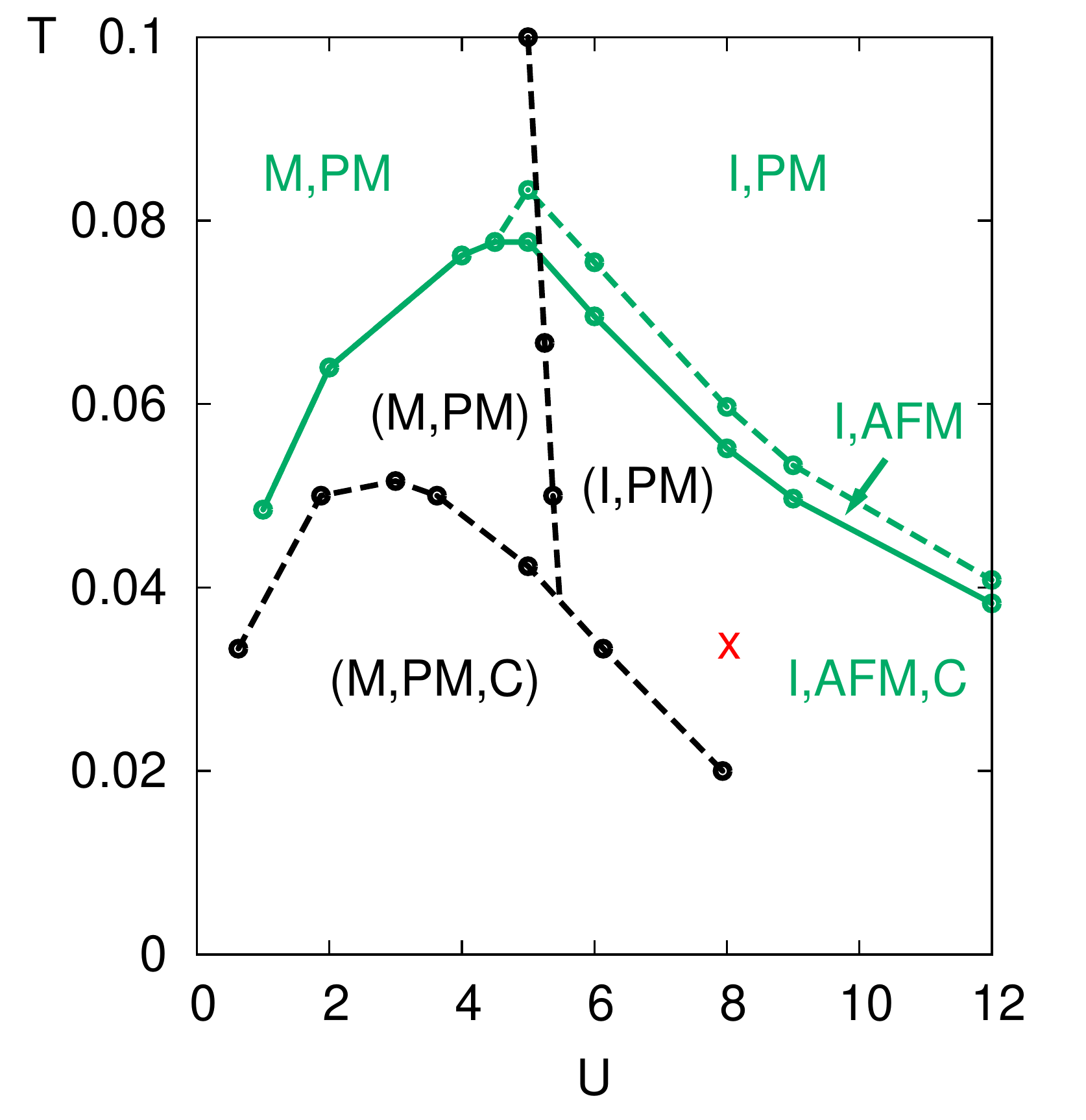} 
\caption{Left panel: Schematic equilibrium and nonequilibrium phase diagrams, and nonthermal trajectory of the photo-doped AFM insulator. The equilibrium phases are labeled as metal (M) or insulator (I), paramagnet (PM) or antiferromagnet (AFM), and according to the presence of composite 
 order (C). Right panel: Equilibrium DMFT+NCA phase diagram (green lines). Black dashed lines indicate the phase diagram restricted to PM states. 
 The red cross locates the initial state considered in the nonequilibrium simulations.  
}
\label{fig_illustration}
\end{center}
\end{figure}

In this paper, we demonstrate the disentangling of a composite (or odd-frequency) orbital order from a magnetic order in a photo-doped nonequilibrium state. Our study is  inspired by the fulleride compounds A$_{3}$C$_{60}$, whose essential physics is captured by  the half-filled three-orbital Hubbard model with negative Hund coupling \cite{Capone2002,Capone2009,Nomura2012}. As sketched in the left panel of Fig.~\ref{fig_illustration}, the high-temperature ($T$) phases of this model break no symmetry, and one observes a crossover or transition from a paramagnetic (PM) metal to a paramagnetic (paired) Mott insulator with increasing interaction. At low $T$, in the vicinity of the Mott transition, a peculiar Jahn-Teller metal with coexisting metallic (M) and Mott insulating (I) orbitals appears \cite{Zadik2015}. This is a composite-ordered (C) phase with a nonzero two-body orbital moment \cite{Hoshino2017,footnote_composite}. 
The low-$T$ Mott state is antiferromagnetic (AFM). We will show that this insulating phase is also of the C type. In equilibrium, the C order is driven by the AFM order, but photo-doping can disentangle the two symmetry breakings. As indicated in Fig.~\ref{fig_illustration}, the photo-induced melting of AFM order results in a hidden gapped PM,C phase, which cannot be stabilized without photo-carriers. 

{\it Model and method.}
The rotationally invariant local Hamiltonian of our three-orbital model reads
\begin{eqnarray}
H_\text{loc} &=& -\mu\sum_{\alpha,\sigma} n_{\alpha\sigma} + U\sum_{\alpha,\sigma} n_{\alpha\sigma}n_{\alpha\bar\sigma} \nonumber\\
&& +\sum_{\alpha>\beta,\sigma} [ (U-2J) n_{\alpha\sigma}n_{\beta\bar\sigma} + (U-3J) n_{\alpha\sigma}n_{\beta\sigma} ]\nonumber\\
&& -J\sum_{\alpha>\beta}(c^\dagger_{\alpha\downarrow}c^\dagger_{\beta\uparrow}c_{\beta\downarrow}c_{\alpha\uparrow} + c^\dagger_{\beta\uparrow}c^\dagger_{\beta\downarrow}c_{\alpha\uparrow}c_{\alpha\downarrow} + \text{h.c.}),\hspace{7mm}
\label{hloc}
\end{eqnarray}
where $\mu$ is the chemical potential, $U$ the intra-orbital repulsion, $J$ the Hund coupling. $c^\dagger_{\alpha\sigma}$ denotes the creation operator for orbital $\alpha$ and spin $\sigma$. In addition, there is a hopping with amplitude $v^{ij}_{\alpha}$ between sites $i$ and $j$, $\sum_{\langle ij\rangle\alpha\sigma}v^{ij}_{\alpha}(t) (c^\dagger_{i\alpha\sigma}c_{j\alpha\sigma} + \text{h.c.})$, which in the real compounds favors different crystallographic axes depending on the orbital \cite{Nomura2012}, enabling orbital-selective excitations. The unique properties of the fulleride compounds originate from the effectively negative $J$ \cite{Capone2009}, which favors low-spin states. We will choose $J=-U/4$, which is larger in magnitude 
than in realistic compounds \cite{Nomura2015}, but does not qualitatively change the physics \cite{Hoshino2017}. 

To compute the nonequilibrium dynamics of the lattice model, we use the nonequilibrium dynamical mean field theory (DMFT) \cite{Georges1996,Freericks2006,Aoki2014}. The lattice problem is mapped to an effective impurity model with a self-consistently determined hybridization function $\Delta_{\alpha\sigma}(t,t')$. Instead of a bcc or fcc lattice, we will consider an infinite-dimensional Bethe lattice, with a semi-circular density of states of bandwidth $4v_{\alpha}(0)$ (in equilibrium at time $t=0$). In this case, the DMFT self-consistency condition directly relates the impurity model Green's function $G_{\alpha\sigma}(t,t')=-i\langle T_\mathcal{C} c_{\alpha\sigma}(t)c^\dagger_{\alpha\sigma}(t')\rangle$ to the hybridization function $\Delta_{\alpha\sigma}$ \cite{Georges1996,Aoki2014}: $\Delta_{\alpha\sigma}(t,t')=v_\alpha(t)G_{\alpha\bar\sigma}(t,t')v_\alpha(t')$, where we switch the spin index between the left and right side in order to describe a potential AFM order.  

To solve the nonequilibrium impurity problem, we use the non-crossing approximation (NCA) \cite{Keiter1971,Eckstein2010}, which captures the essential physics in the strongly correlated regime. To mimic the effect of a photo-excitation, we will apply a hopping modulation of the form
\begin{equation}
v_\alpha(t)=v_\alpha(0)+a_\alpha f(t-t_p)\sin(\Omega(t-t_p))
\end{equation}
with $a$ the amplitude of the modulation, $f(t-t_p)=\exp[(-0.3(t-t_p))^2]$ a Gaussian envelope function centered at time $t_p$, and $\Omega$ the frequency of the modulation. For $\Omega$ larger than the gap, this creates mobile charge carriers, and thus has an effect similar to photo-doping. We use $v(0)$ as the unit of energy ($\hbar/v(0)$ as the unit of time) and set $\hbar=1$.  

The C phases have no conventional orbital order, but are characterized by an orbital-dependent double occupation, or a nonzero 
\begin{equation} 
T^8_\text{comp}(t)=\sum_\alpha \sqrt{3}\lambda_{\alpha\alpha}^8 \langle n_{\alpha\uparrow}(t)n_{\alpha\downarrow}(t) \rangle ,
\end{equation}
with $\lambda^8=\frac{1}{\sqrt{3}}\text{diag}(1,1,-2)$ a Gell-Mann matrix. 
This symmetry breaking can also be regarded as an odd-frequency orbital-order \cite{Hoshino2017} and detected through the time-dependence of the Green's functions.
In equilibrium, the imaginary-time dependent quantity
$T^8(\tau)=\sum_{\alpha\sigma}\sqrt{3}\lambda^8_{\alpha\alpha}\langle c^\dagger_{\alpha\sigma}c_{\alpha\sigma}(\tau)\rangle$ can be expanded as 
$T^8(\tau)=T^8_\text{even}+T^8_\text{odd}\tau + O(\tau^2)$
to define the order parameters for conventional orbital order ($T^8_\text{even}$) and for odd-frequency orbital order ($T^8_\text{odd}$). 
Since $T^8_\text{comp}$  appears in the expression for $T^8_\text{odd}$~\cite{Hoshino2017}, they can both be used to detect the C order. 
For nonequilibrium calculations, it is convenient to express $T^8_\text{odd}$ as 
\begin{equation}
T^8_\text{odd}(t)=-\sum_{\alpha\sigma}\sqrt{3}\lambda^8_{\alpha\alpha}\partial_{t'} G_{\alpha\sigma}^<(t',t)\Big|_{t'=t}. 
\end{equation}
We also consider AFM order, $\text{AFM}(t)=\sum_\alpha \langle n_{\alpha\uparrow}(t) -  n_{\alpha\downarrow}(t)\rangle$, while antiferro-orbital order and superconductivity, which also exist in this model \cite{Ishigaki2019}, are suppressed. 

{\it Results.}
The right panel of Fig.~\ref{fig_illustration} 
shows the DMFT+NCA phase diagram of the half-filled model in the space of $U$ and $T$. Within NCA, the metallic, paramagnetic, composite-ordered M,PM,C phase and low-$T$ metal-insulator crossover line of the PM calculation are buried inside an extended insulating composite-ordered I,AFM,C solution (solid green line), which features one AFM and two almost PM orbitals. On the strong coupling side, there is a narrow range of conventional AFM insulator with three degenerate AFM orbitals, close to $T_\text{N\'eel}$. The transition from the high-$T$ PM to the I,AFM phase is second order, while the transition to I,AFM,C is first order. In the exact DMFT solution (obtained with a CT-HYB impurity solver \cite{Gull2011}), 
only the paired Mott region becomes AFM \cite{Hoshino2017}, while the metallic and composite-ordered phases at smaller $U$ are PM. We checked that this exact AFM solution is also of the C type, with a direct first order transition from the I,PM phase to the I,AFM,C solution. In the nonequilibrium analysis, we will restrict ourselves to $U=8$, and initial $T=0.033$ (red cross in Fig.~\ref{fig_illustration}). In this large-$U$ regime, apart from the detailed behavior near the phase boundary, the physics is correctly captured by the NCA solver.

\begin{figure}[t]
\begin{center}
\includegraphics[angle=0, width=0.8\columnwidth]{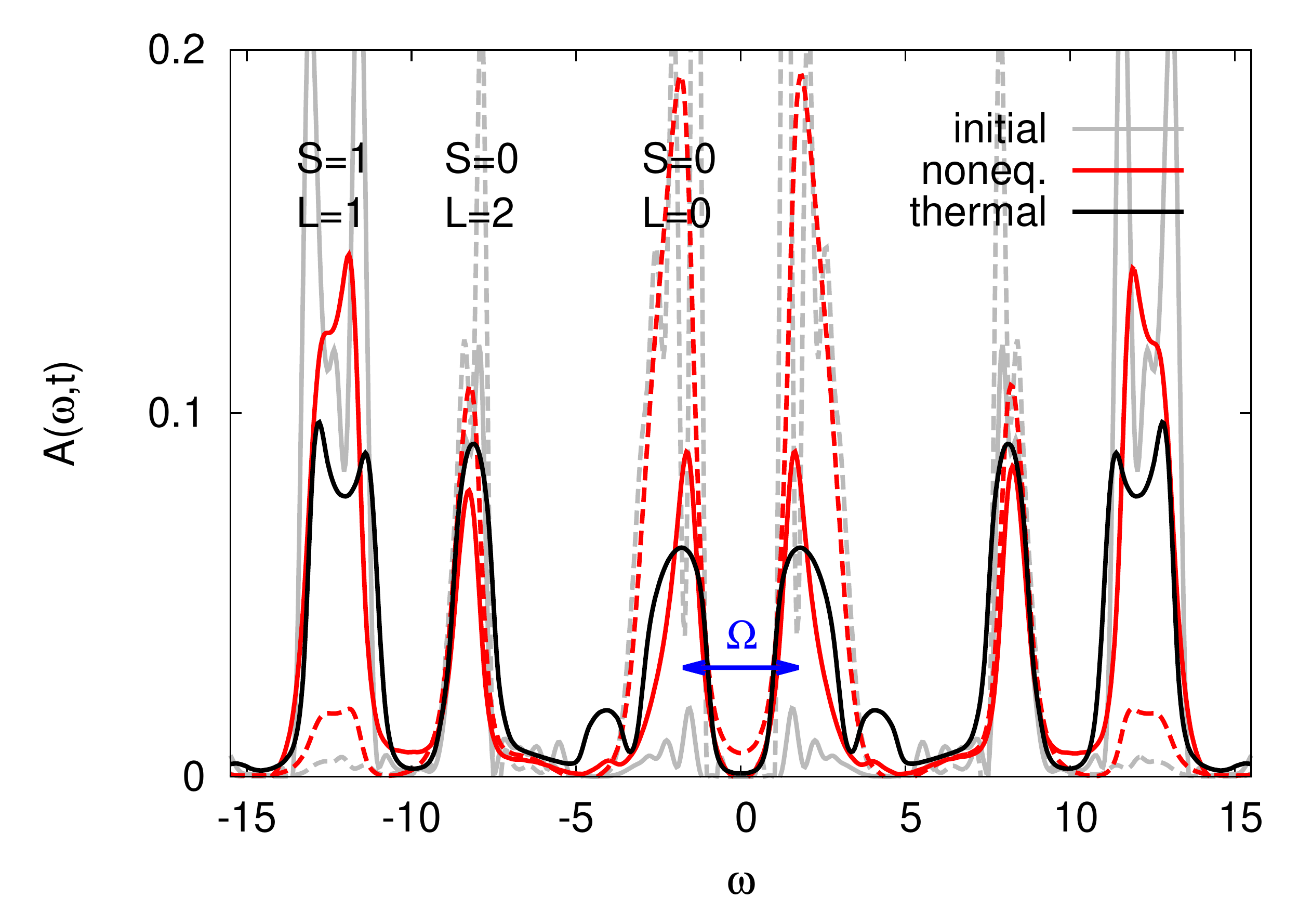} 
\includegraphics[angle=0, width=0.8\columnwidth]{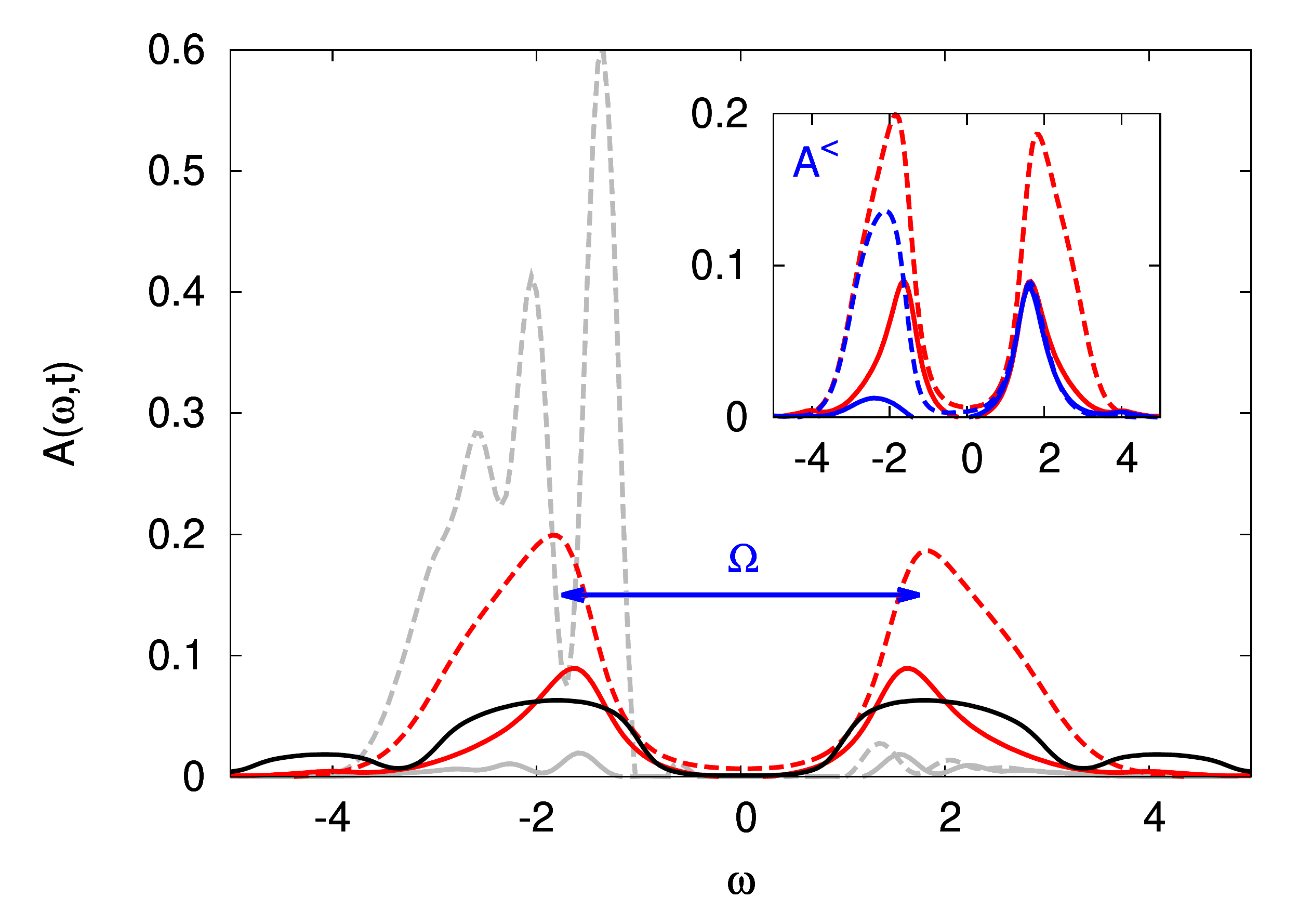} 
\caption{Top panel: Spectral functions in the initial I,AFM,C state at $U=8,T=0.033$, the metastable I,PM,C state (measured at $t=10$), and the thermalized I,PM state. In the AFM states, solid (dashed) lines indicate the spin-averaged spectra for orbital $\alpha=1,2$ ($\alpha=3$). We also indicate the 
total $S$ and $L$ associated with the different peaks. Bottom panel: Majority-spin spectral functions. 
The blue lines in the inset show the occupied density of states $A^<(\omega,t=10)$ for orbital $\alpha=1,2$ (solid line) and $\alpha=3$ (dashed line).
}
\label{fig_spectra}
\end{center}
\end{figure}

\begin{figure*}[ht]
\begin{center}
\includegraphics[angle=0, width=0.64\columnwidth]
{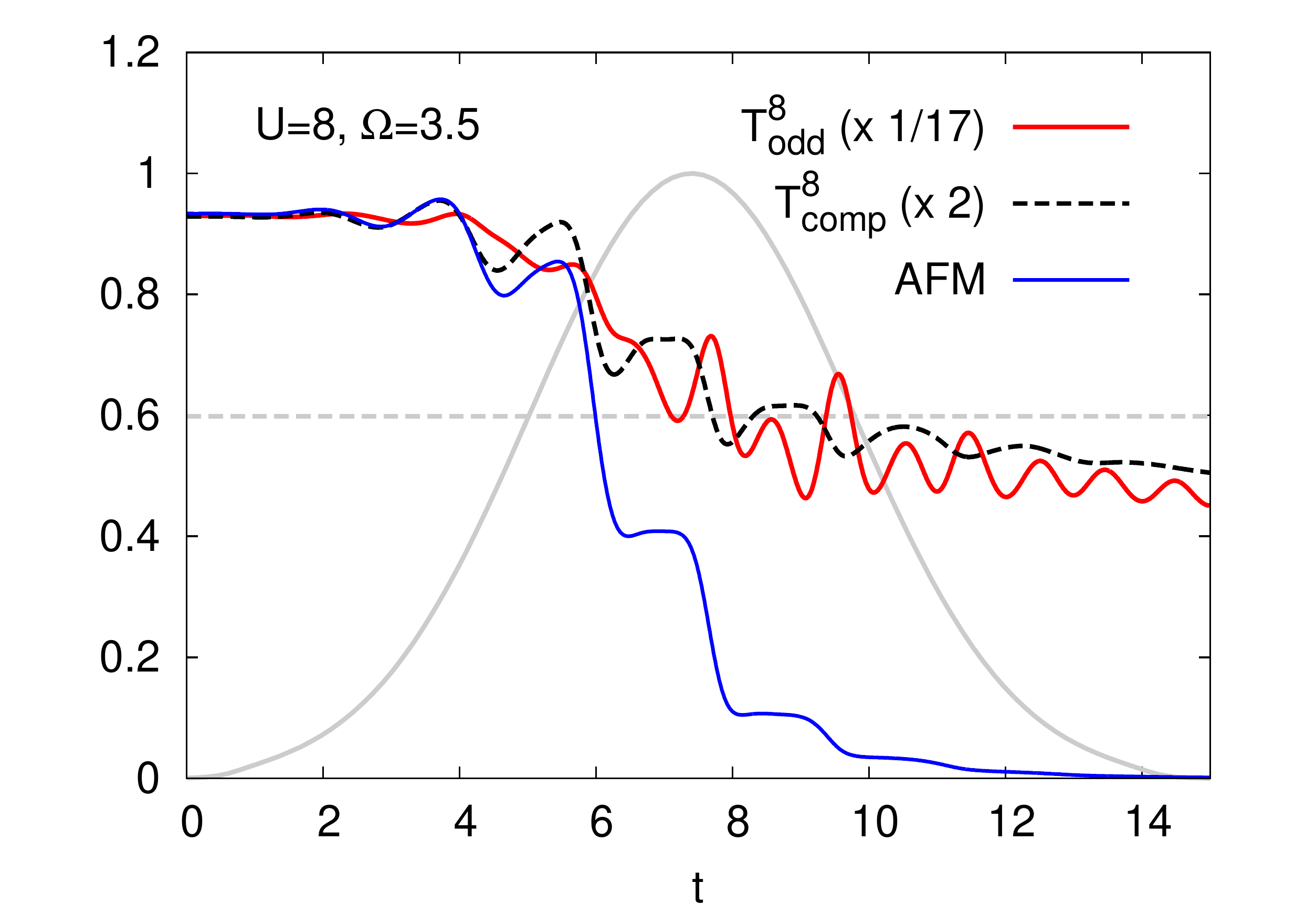} 
\hfill
\includegraphics[angle=0, width=0.64\columnwidth]{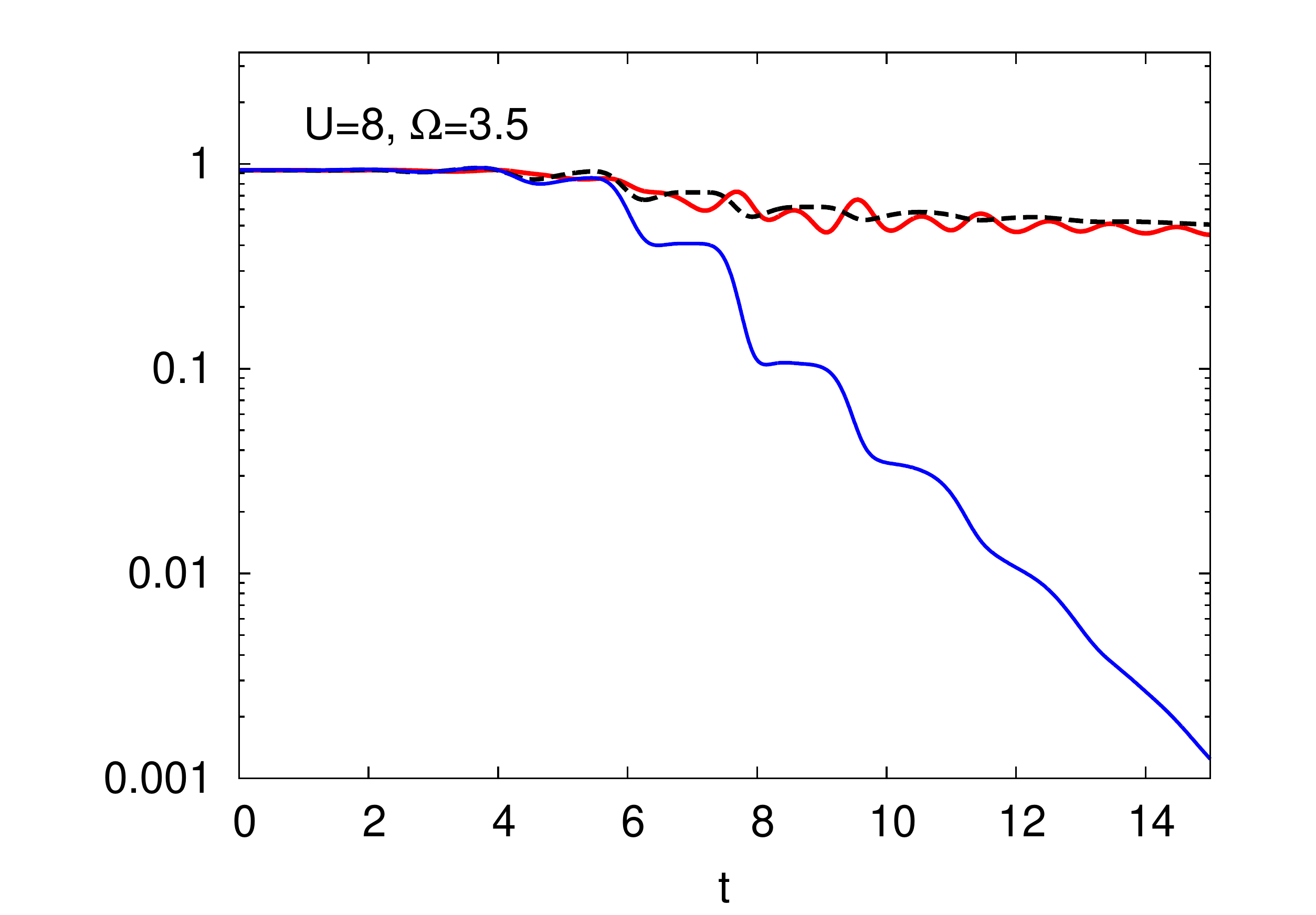} 
\hfill
\includegraphics[angle=0, width=0.66\columnwidth]{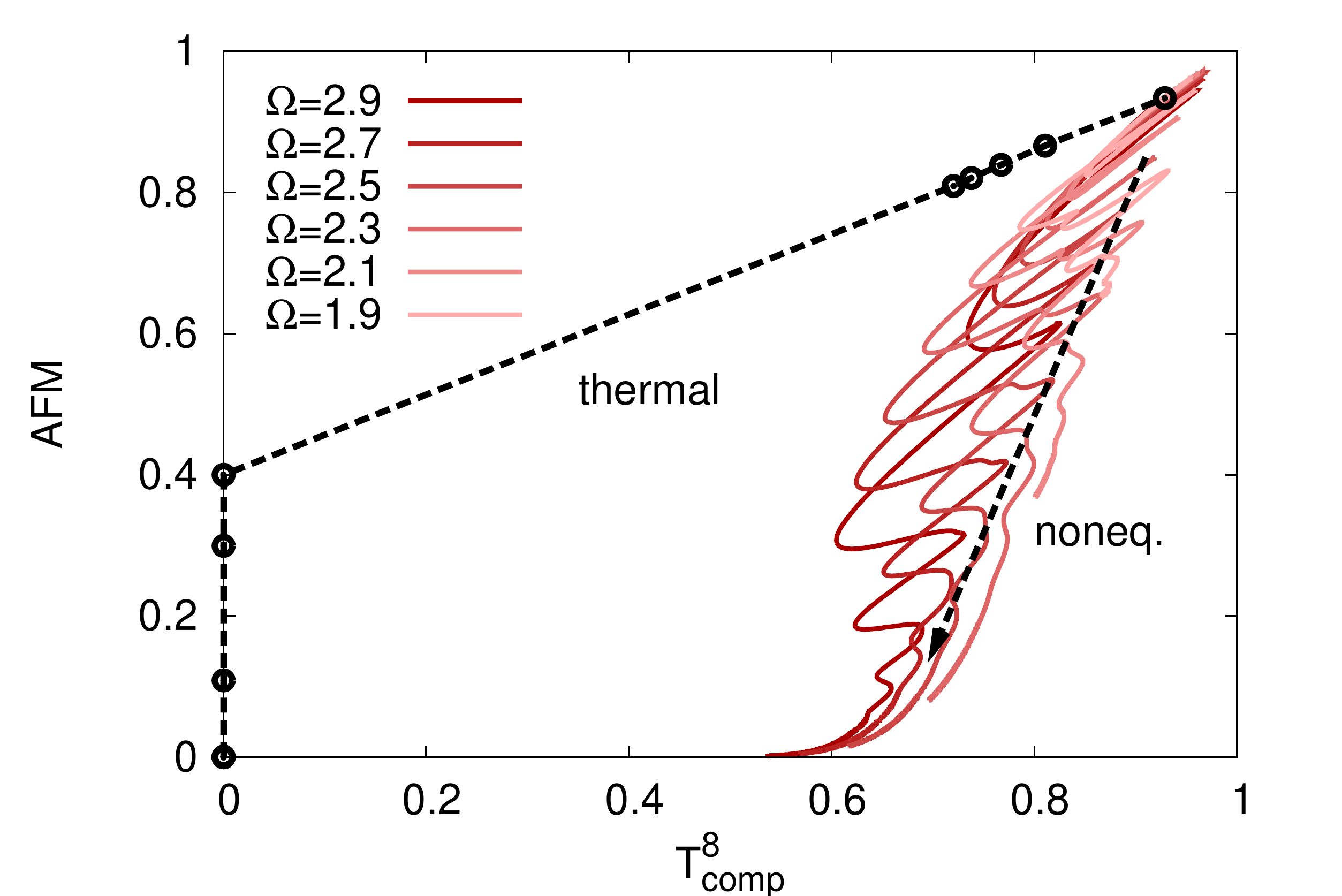} 
\caption{Left two panels: Evolution of the AFM, $T^8_\text{odd}$ and $T^8_\text{comp}$ order parameters for the photo-doped I,AFM,C state with $U=8$ and initial $T=0.033$ ($\Omega=3.5$, $a=1$).   
The solid grey line indicates the envelope $f(t-t_p)$ of the pulse, and the horizontal dashed line the expected reduction of  $T^8_\text{comp}$ due to the photo-doped population of doublons and holons. 
Right panel:  AFM versus composite order after pulses with indicated frequencies $\Omega$ and $a=1$. The black dashed line with circles shows the order parameters for different $T$ in equilibrium.  
All pulses are applied to orbital 3.
}
\label{fig_afm_u8}
\end{center}
\end{figure*}

The nature of the I,AFM,C state is revealed by the orbital-resolved local spectral function $A_{\alpha\sigma}(\omega,t)=-\frac{1}{\pi}\int_t^{t_\text{max}} dt' e^{i\omega(t'-t)}G_{\alpha\sigma}^R(t',t)$, which for the initial equilibrium state is shown by the grey lines in the upper panel of Fig.~\ref{fig_spectra}.  Two orbitals ($\alpha=1,2$ in our simulations, solid lines) are PM and in a paired Mott insulating state, which is stabilized by the pair-hopping term. 
The third orbital ($\alpha=3$, dashed line) is in an AFM insulating state. (For simplicity, we plot the spin-averaged spectrum.) 
Because the two types of orbitals have different double occupations and different Green's functions, $T^8_\text{comp}$ and $T^8_\text{odd}$ are nonzero (while $T^8_\text{even}=0$). 
The spectral function features three pairs of peaks, with energy separations $\Delta\omega\approx U+2J=4$, $U-4J=16$, $U-8J=24$. These 
correspond to electron removal/addition from/to the lowest energy half-filled states with total spin $S=\tfrac{1}{2}$ and total angular momentum $L=1$ 
(see e.~g. Tab.~1 in Ref.~\onlinecite{Georges2013}). The lowest peaks are associated with the creation of four- and two-electron states with $S=0, L=0$,  
the middle peaks with $S=0,L=2$, and the high-energy peaks with $S=1,L=1$.  

The half-filled $S=\tfrac12,L=1$ atomic states are superpositions of the states $|dh\rangle_{\alpha\beta}\otimes|s\rangle_{\gamma}$, where $|dh\rangle_{\alpha\beta}=\frac{1}{\sqrt{2}}({|\!\uparrow\downarrow,0\rangle_{\alpha\beta}}+|0,\uparrow\downarrow\rangle_{\alpha\beta})$ describes a doublon-holon eigenstate of the pair-hopping term (eigenenergy $J$) and $|s\rangle_{\gamma}=|\sigma\rangle$ a singly occupied orbital ($\alpha, \beta, \gamma$ is a cyclic permutation of $1,2,3$). Adding/removing an electron to/from $|s\rangle_{\gamma}$ creates a superposition of $S=0,L=0$ and $S=0,L=2$ states, while in the case of $|dh\rangle_{\alpha\beta}$, one ends up in a superposition of $S=0,L=2$ and $S=1,L=1$ states. 

In the I,AMF,C phase, the orbital symmetry is broken, and our simulation produces an insulator dominated by $|dh\rangle_{12}\otimes|s\rangle_{3}$. The above considerations explain why the spectral function for orbital $\alpha=1,2$ (solid gray line) has weight mainly in the middle and high energy peaks, while $\alpha=3$ (dashed gray line) contributes to the middle and low energy peaks. 
The lower panel of Fig.~\ref{fig_spectra} shows the majority-spin spectral functions and reveals that the AFM order is supported by orbital $3$, with a large spin polarization and characteristic spin-polaron peaks \cite{Werner2012,Taranto2012}.
We can understand the origin of the orbital symmetry breaking in the lattice system by considering the effect of inter-site hopping. 
Because of the orbital-diagonal hopping, the $|dh\rangle_{12}\otimes|s\rangle_{3}$ arrangement with AFM ordered $|s\rangle_{3}$ fully exploits the spin-exchange energy. In an orbitally symmetric system, most of the virtual hoppings are agnostic to the spin arrangement and do not stabilize an AFM order. Hence, the composite order in the equilibrium I,AFM,C phase is driven by the magnetic order.     

We now show that the C order can be disentangled from the AFM order by photo-doping. 
We apply a pulse with $\Omega=3.5$ and $a=1$ to orbital $\alpha=3$ (see supplementary material (SM) for the orbital-symmetric pulse) 
and track the evolution of the AFM, $T^8_\text{odd}$ and $T^8_\text{comp}$ order parameters. The left two panels of Fig.~\ref{fig_afm_u8} show the results on a linear and logarithmic scale. This pulse with energy comparable to the splitting between the $S=0, L=0$ multiplets 
quickly melts the AFM order, while $T^8_\text{odd}$ and $T^8_\text{comp}$ (after some suppression during the pulse) decrease much more slowly. 
The result is a transient I,PM,C phase, which has pure odd-frequency or C order. 
Such a symmetry-broken state cannot be realized in equilibrium, where the I,PM  phase has degenerate orbitals.  

The energy gap and orbital symmetry breaking in the hidden I,PM,C phase are clearly revealed by the red spectra for $t=10$ in Fig.~\ref{fig_spectra}, where the solid (dashed) lines show the results for $\alpha=1,2$ ($\alpha=3$), and in the top panel we average over spin. 
In contrast, the equilibrium M,PM,C phase, which exists at smaller $U$, is an orbital-selective metal state at elevated $T$, with a large but orbital-dependent density of states at $\omega=0$ \cite{Yue2020b}. The nonthermal nature of the red spectra becomes obvious by comparison to the spin and orbital symmetric spectra of the thermalized system expected after a long time (black line). The corresponding $T=1.42$ has been determined from the conserved total energy after the pulse.

The thermalization pathway is illustrated in the right panel of Fig.~\ref{fig_afm_u8}, which plots the evolution of the system for different pulse frequencies $\Omega$ in the space of the AFM and $T^8_\text{comp}$ order parameters. In equilibrium, as $T$ is increased, the AFM order decreases more slowly than the composite order, then (within DMFT+NCA) a jump to the conventional I,AFM state with $T^8_\text{comp}=0$ occurs, and finally this I,AFM order melts (black dashed line with circles). The nonthermal evolution follows a distinctly different pathway, as illustrated by the dashed arrow. In this case, the AFM  order is quickly suppressed, while the C order persists, so that the system switches into the hidden I,PM,C phase. A qualitatively similar behavior, but for conventional orbital and magnetic orders, has been observed in a photo-doped two-orbital model in Ref.~\cite{Li2018}.  

\begin{figure}[t]
\begin{center}
\includegraphics[angle=0, width=0.85\columnwidth]{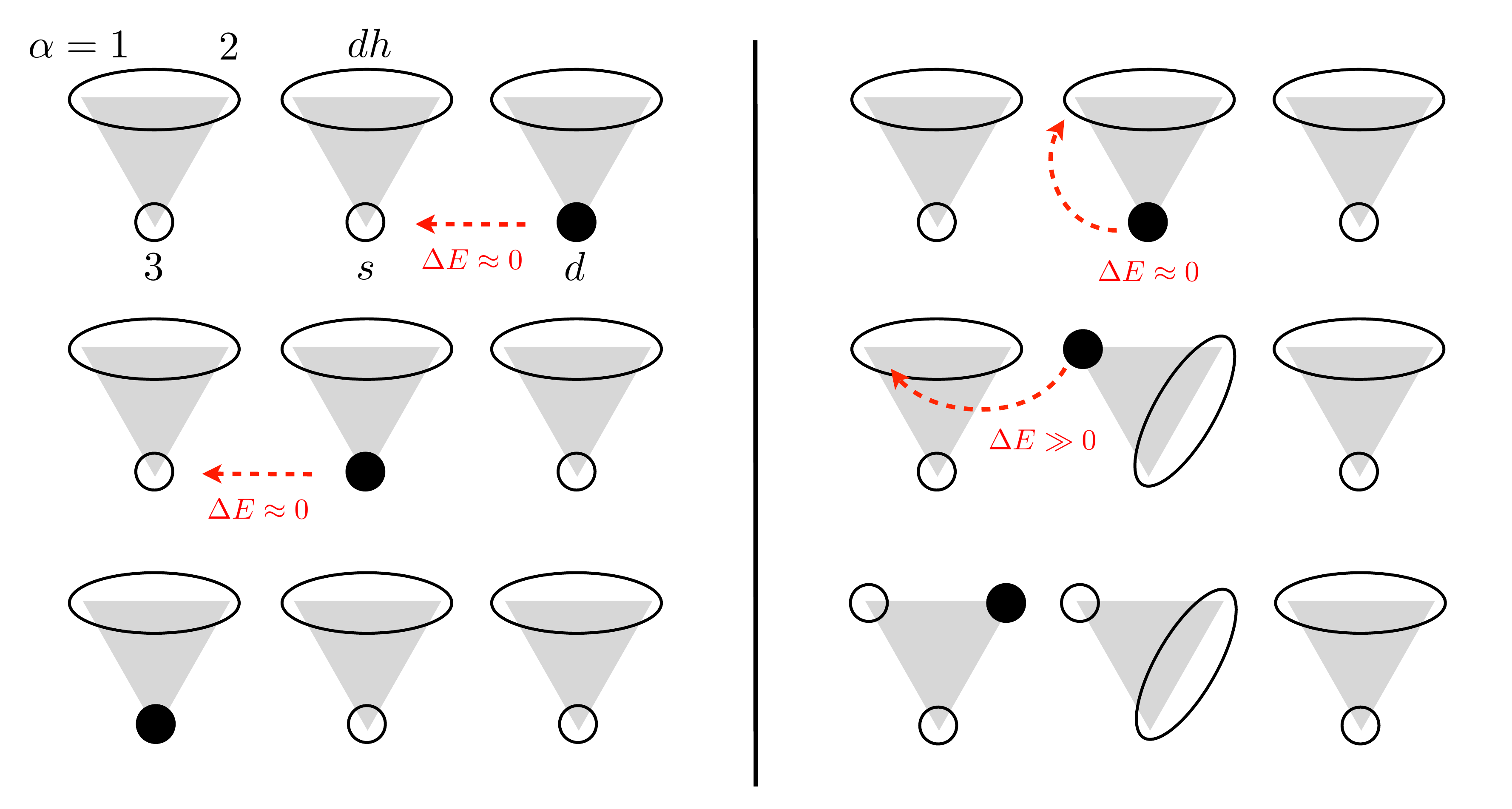} 
\caption{Left panels: Illustration of doublon hopping in orbital $\alpha=3$ in the I,PM,C state. Full/empty dots represent doubly/singly occupied sites in orbital 3 and the black oval the doublon/holon state $|dh\rangle_{12}$. Right panels: Illustration of a doublon hopping process involving orbitals $\alpha=1,2$. In the first step, pair hopping leads to the reshuffling $|dh\rangle_{12}\otimes |d\rangle_3 \rightarrow |dh\rangle_{23}\otimes |d\rangle_1$. In the last step, the doublon in orbital $1$ hops to the neighboring $|dh\rangle_{12}$ doublon-holon state.  
}
\label{fig_hopping}
\end{center}
\end{figure}

The selective melting of the AFM order and the emergence of the I,PM,C state are linked to the kinematics of the excited doublons and holons. 
 The photo-doping pulse with $\Omega=3.5$ produces them in the lowest-energy $S=0,L=0$ multiplet and hence orbital $\alpha=3$ (see blue arrows in Fig.~\ref{fig_spectra}).
 Since the gap size is comparable to the width of the $S=0,L=0$ subband, the injected doublons/holons are long-lived \cite{Sensarma2010,Eckstein2011} and as illustrated in the left panels of Fig.~\ref{fig_hopping}, they can move to neighboring singly occupied orbitals 3 without any local energy cost.
 This hopping disturbs the AFM spin background and results in the melting of the magnetic order, because the initial kinetic energy of the doublons/holons is larger than the spin exchange energy \cite{footnote_splitting} and their density is sufficiently high. 
The electron-spin interactions and the thermalization bottleneck lead to a PM state in which much of the injected energy is stored as potential energy, and the doublons/holons are relatively cold \cite{Werner2018}. Such nonequilibrium quasi-steady states can host nonthermal electronic orders \cite{Murakami2021,Li2020,Li2021}.

Fitting the nonequilibrium distribution function $A^<(\omega,t)/A(\omega,t)$ to a Fermi function yields $T_\text{eff}=0.38$ for the doublons/holons. This is substantially lower than the $T=1.42$ of the thermalized state, but 
still high compared to the temperature scales of the equilibrium C phases. Entropy cooling \cite{Werner2019b} or the large susceptibility to orbital symmetry breaking in low-temperature states (see SM) can thus not fully explain the robustness of the hidden I,PM,C phase.  Rather, the C order survives because the orbital symmetry-breaking allows the doublons/holons to move around and to optimize the kinetic energy. 
This mechanism is similar to the stabilization of ferromagnetic order by itinerant carriers in the double exchange model \cite{Zener1951,deGennes1960}, but here we are dealing with nonthermal carriers and a ferro-type two-body orbital moment. 

To support this interpretation, we consider the dynamics of the injected doublons.  In principle, as illustrated in the right panels of Fig.~\ref{fig_hopping}, the pair-hopping term can reshuffle a configuration with additional doublon, and for example convert $|dh\rangle_{12}\otimes |d\rangle_3$ into $|dh\rangle_{23}\otimes |d\rangle_1$. However, in the C phase with broken orbital symmetry, an electron from orbital $\alpha=1$ cannot be easily transferred to the neighboring $|dh\rangle_{12}$ state, since such a process corresponds to electron insertion into the intermediate- or high-energy multiplets. The corresponding substantial energy cost 
in most cases prevents the hopping, so that the local configuration reverts to $|dh\rangle_{12}\otimes |d\rangle_3$ and the doublon moves on in orbital $3$, leaving behind an unperturbed I,PM,C pattern. 

Since local states with photo-doped doublons and holons can be reshuffled by the pair hopping term, they should not contribute to $T^8_\text{comp}$. The pulse used in the left panel of Fig.~\ref{fig_afm_u8} produces doublons or holons on $2\times 18=36\%$ of the sites. The horizontal grey dashed line shows the initial $T^8_\text{comp}$ reduced by these $36\%$ and confirms that the suppression of the C order during the pulse is due to the photo-doped carriers, while the orbital background remains undisturbed. A further consequence of the local reshuffling of doublons is the fact that the occupied density of states in the upper Hubbard band is almost independent of orbital (see inset of Fig.~\ref{fig_spectra}). The small difference between the occupation $A^<$ (blue) and the spectral function $A$ (red) for $\alpha=1,2$ indicates that doublons in these orbitals are almost exclusively generated by such reshufflings, while for $\alpha=3$ only a fraction of the states are occupied, consistent with the picture of itinerant doublons moving in a Hubbard band (Fig.~\ref{fig_hopping}). 

{\it Conclusions.}
We have shown that the AFM insulating state in a simple model for A$_3$C$_{60}$ exhibits a spontaneous symmetry breaking into two 
paired Mott insulating orbitals and a third conventional Mott insulating orbital which supports the AFM order.  The photo-induced nonthermal melting of the AFM order converts this equilibrium I,AFM,C phase into a hidden I,PM,C phase with pure composite or odd-frequency orbital order (Fig.~\ref{fig_illustration}). 
This hidden state is stabilized by the long life-time and orbital-dependent kinetic energy of the photo-carriers. It would thus be an interesting target for studies with nonequilibrium steady state \cite{Li2021} or effective equilibrium \cite{Murakami2021} formalisms. While $|J|$ is much smaller in realistic fulleride compounds, signatures of such a hidden state may be detectable in photo-doped bcc Cs$_3$C$_{60}$.  

{\it Acknowledgements.} 
The calculations have been performed on the Beo04 and Beo05 clusters at the University of Fribourg, using a code based on the NESSi library \cite{Nessi}. PW acknowledges support from ERC Consolidator Grant No.~724103 and SNSF Grant No.~200021\_196966. YM acknowledges support from a Grant-in-Aid for Scientific Research from JSPS, KAKENHI Grant Nos.~JP20K14412, JP20H05265 and JST CREST Grant No.~JPMJCR1901.

\clearpage 
 
\appendix

\begin{center}
{\bf Supplemental Material}
\end{center}

\section{Symmetric pulse}

\begin{figure}[b]
\begin{center}
\includegraphics[angle=0, width=0.49\columnwidth]{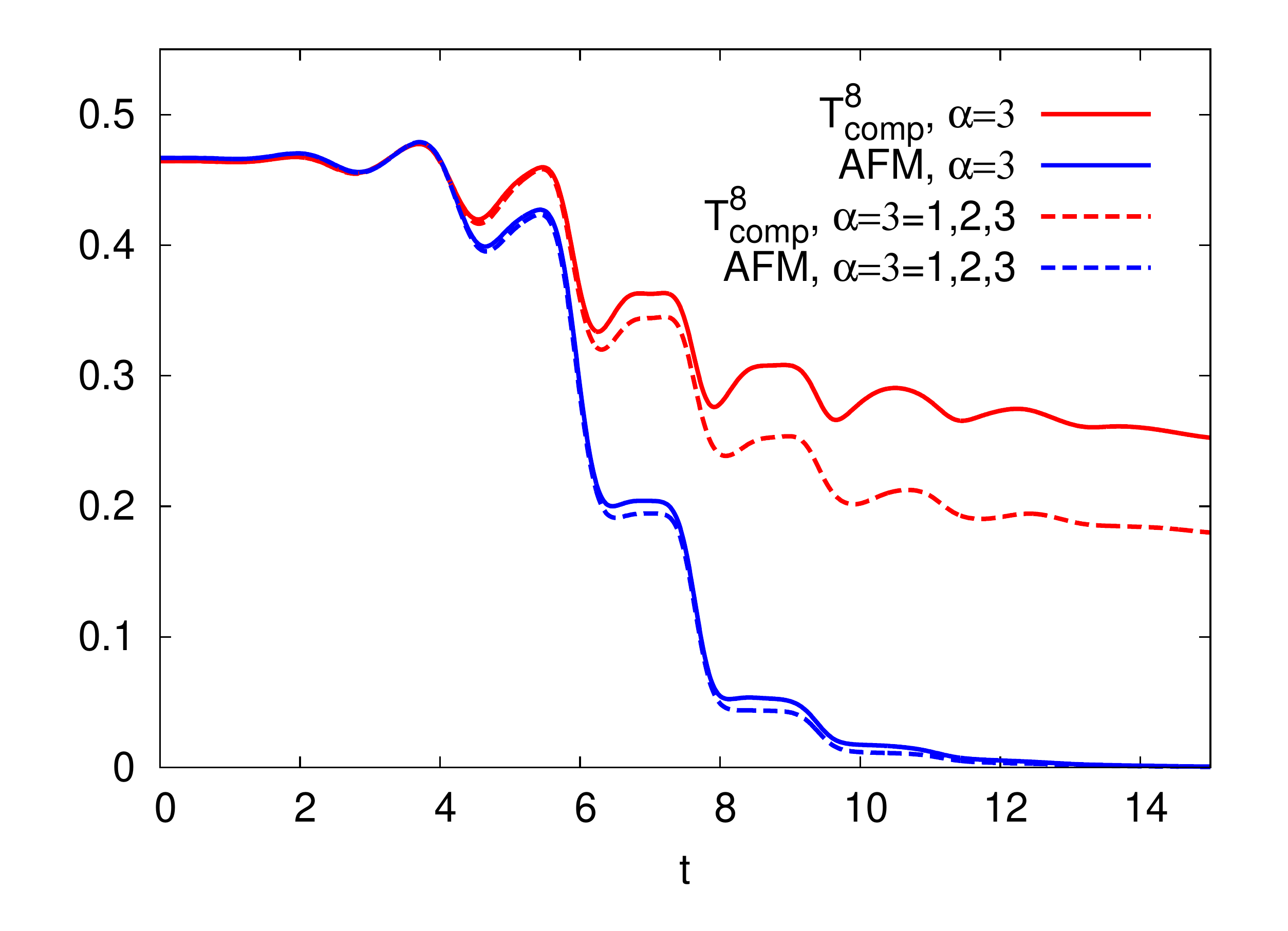} 
\hfill
\includegraphics[angle=0, width=0.49\columnwidth]{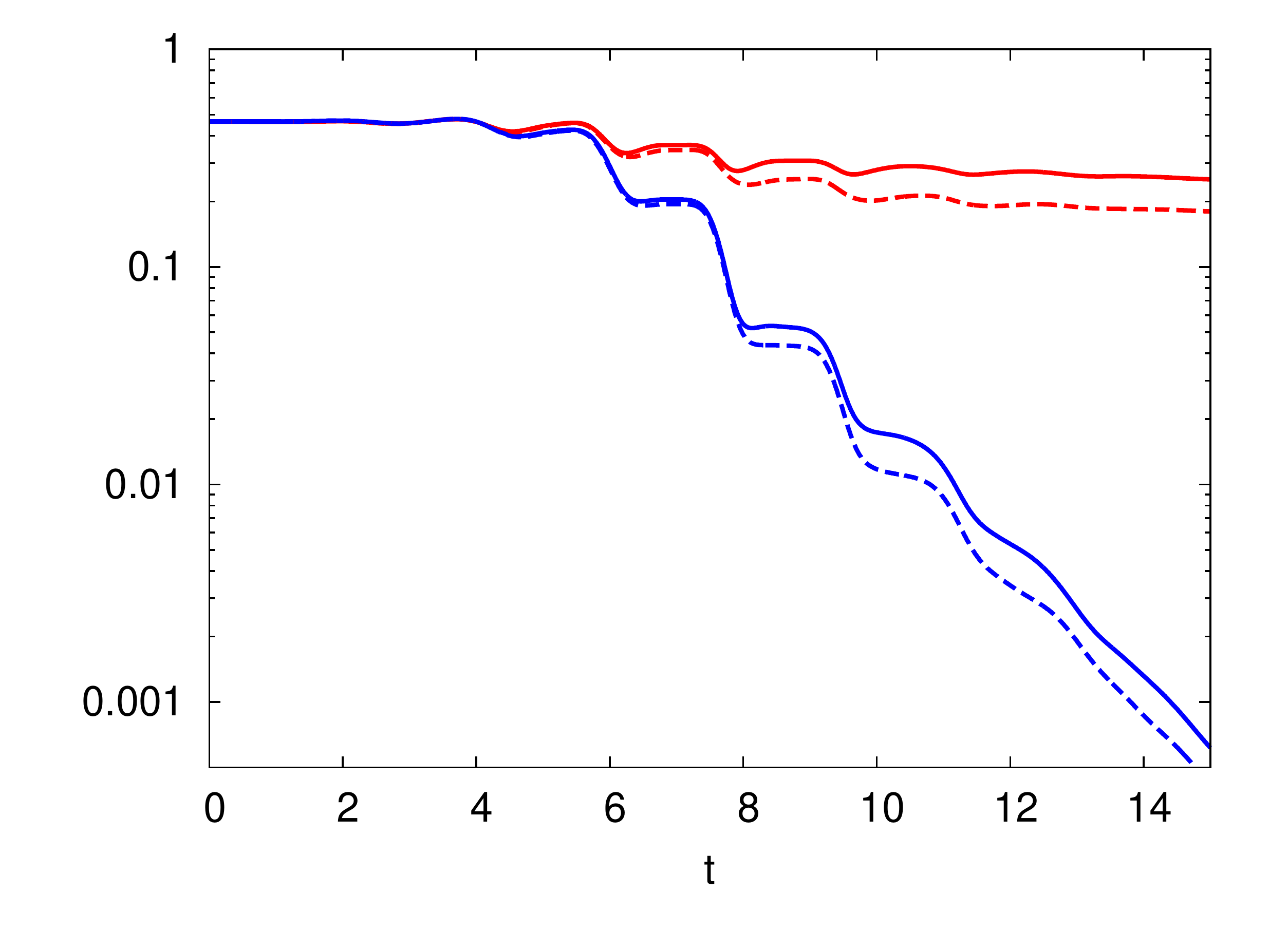} 
\caption{Time evolution of the $T^8_\text{comp}$ and AMF orders after a photo-excitation, plotted on a linear (left) and logarithmic (right) scale. Solid lines show simulations for a pulse applied to orbital $\alpha=3$ and dashed lines for a pulse applied to all three orbitals. The parameters are the same as in the main text ($U=8$, $\Omega=3.5$, $a=1$). 
}
\label{fig_sym}
\end{center}
\end{figure}

Figure~\ref{fig_sym} shows the evolution of the $T^8_\text{comp}$ and AFM order parameters in a photo-doped system with $U=8$, $\Omega=3.5$, $a=1$ for two types of excitations: (i) pulse applied only to orbital 3 (solid lines, same as in Fig.~\ref{fig_afm_u8} of the main text), and (ii) pulse applied to all three orbitals (dashed lines). While $T^8_\text{comp}$ is more strongly suppressed by the second pulse, because of the larger density of doublons and holons produced, the result remains qualitatively the same. The photo-doping rapidly melts the AMF order, while the composite order survives for much longer times, resulting in a metastable I,PM,C state. 

Since orbitals $\alpha=1,2$ have little spectral weight near $\omega = \pm \tfrac{\Omega}{2}$ (see Fig.~\ref{fig_spectra} of the main text), most doublons and holons are produced in orbital $\alpha=3$, even if the pulse is applied to all orbitals. 

\section{Local density matrix}

\begin{figure}[t]
\begin{center}
\includegraphics[angle=0, width=0.49\columnwidth]{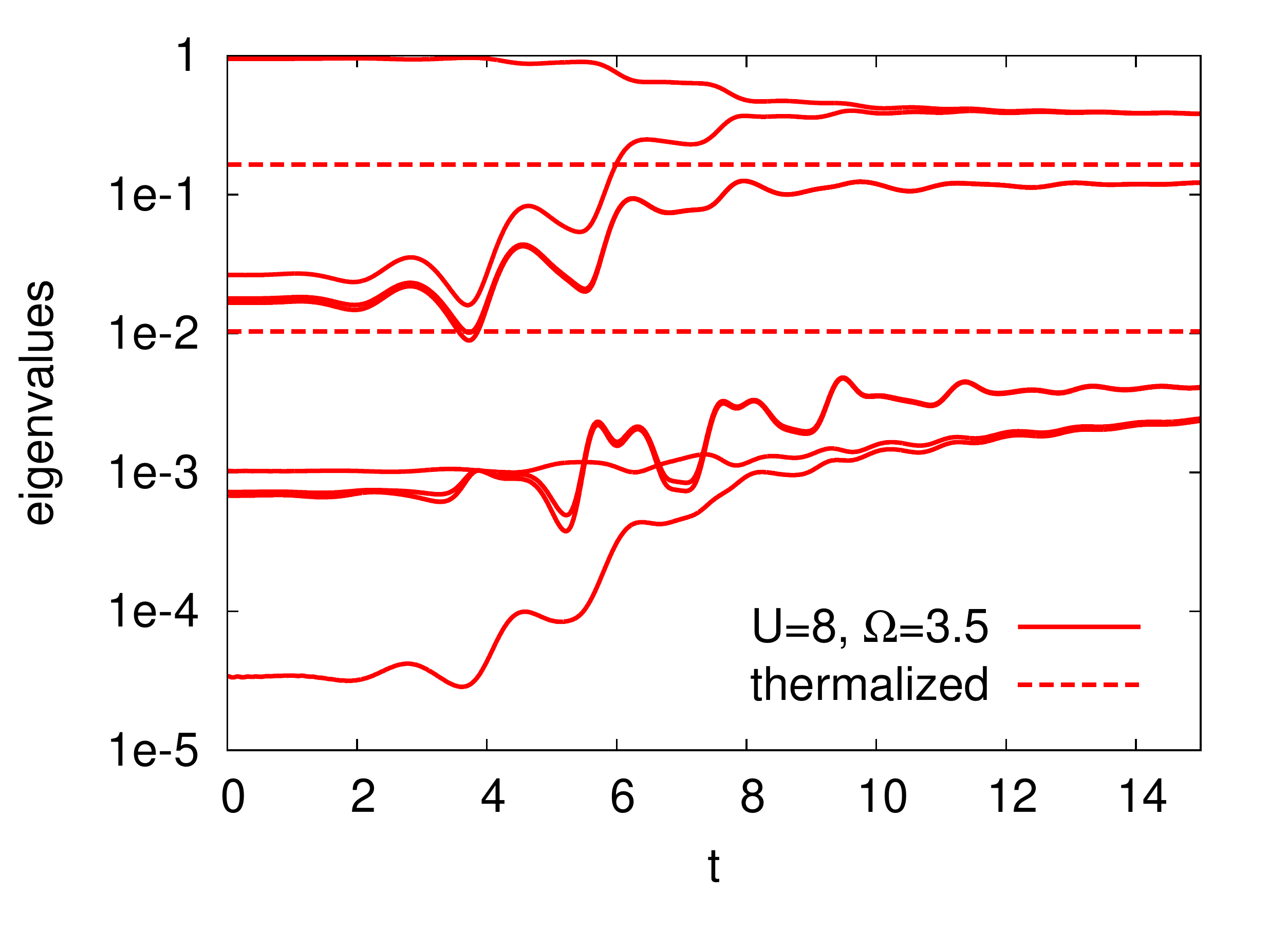} 
\hfill
\includegraphics[angle=0, width=0.49\columnwidth]{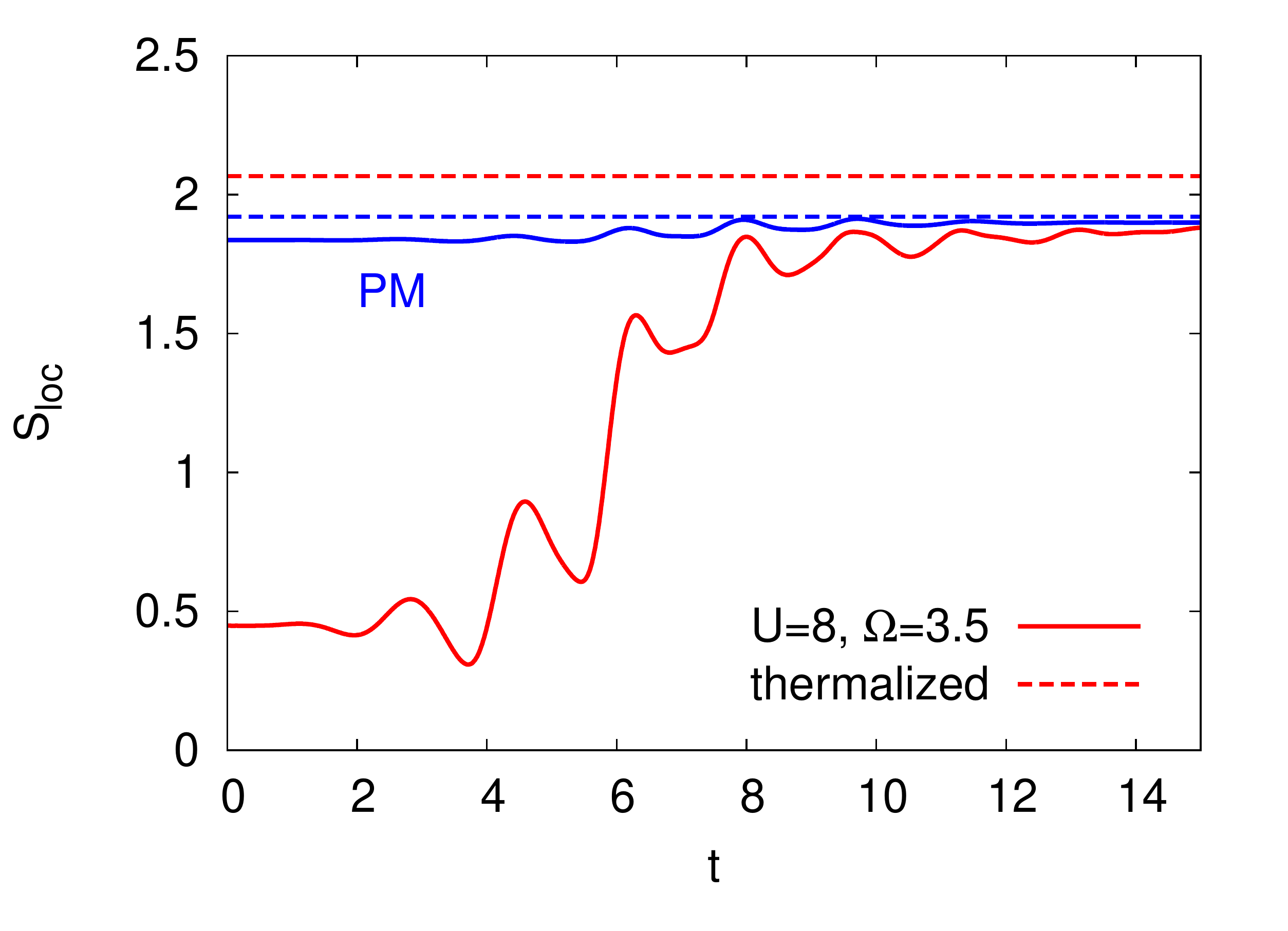} 
\caption{Photo-doped I,AFM,C state with $U=8$ and initial $T=0.033$ ($\Omega=3.5$, $a=1$). The left panel shows the evolution of the eigenvalues $e_1$,\dots,$e_{12}$ and the right panel the evolution of the local entropy $S_\text{loc}$. Red dashed lines indicate the values reached in the thermalized state with $T=1.42$. In the right panel, we also show the analogous results for simulations starting from the I,PM state (blue lines, thermalized state with $T=0.92$). 
}
\label{fig_entropy}
\end{center}
\end{figure}

It is also instructive to explore the local density matrix and local entropy of the photo-doped system \cite{Werner2020}. We have measured $\rho_{\mu\nu}=|\mu\rangle\langle\nu|$ in the space of half-filled states with one empty, one singly-occupied and one doubly-occupied orbital ($12\times 12$ matrix). The left panel of Fig.~\ref{fig_entropy} plots the evolution of the eigenvalues $e_1,\ldots,e_{12}$ of $\rho$ after the photo-doping pulse and the right panel plots the local entropy defined as $S_\text{loc}=-e_1\ln _1-e_2\ln _2\ldots -e_{12}\ln e_{12}$.   The corresponding results for the thermalized state at $T=1.42$ are indicated by the red dashed lines. 

In the initial I,AFM,C state, a single eigenvalue close to 1 dominates, because of the unique dominant $|dh\rangle_{12}\otimes|s\rangle_{3}$ state, and this results in a low $S_\text{loc}$. In the thermalized state without symmetry breaking, there are two groups of six eigenvalues (dashed lines), associated with ``bonding" and ``anti-bonding" combinations of doublons and holons. The hidden state has a distinctly different eigenvalue structure. Here, the largest eigenvalue becomes doubly degenerate, because of the spin symmetry in $|s\rangle_\gamma$, while the orbital symmetry breaking results in a gap to the eigenvalues of the four sub-dominant states and a reduced $S_\text{loc}$.

\section{Susceptibility for orbital symmetry breaking}

\begin{figure}[b]
\begin{center}
\includegraphics[angle=0, width=0.5\columnwidth]{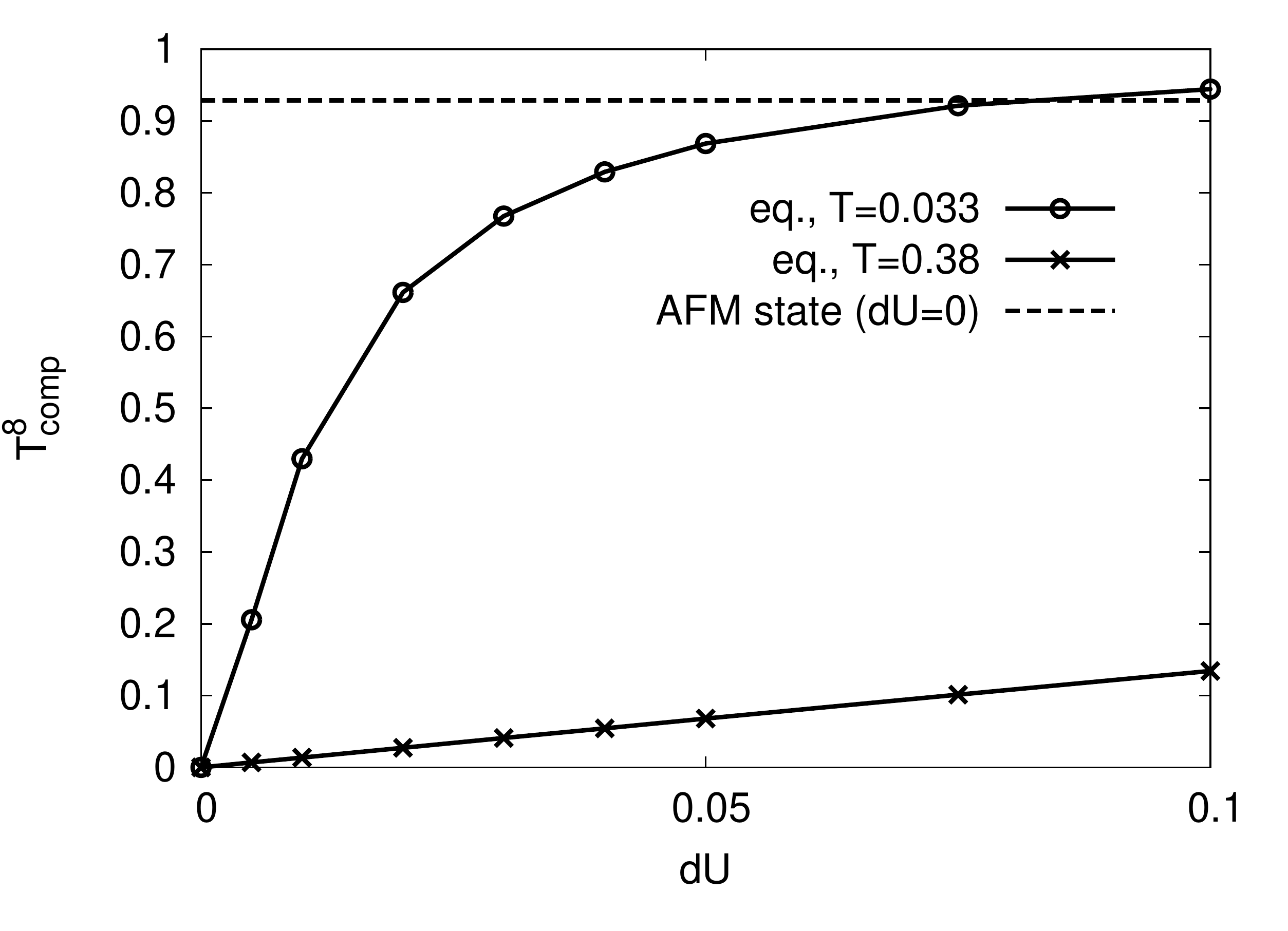} 
\caption{$T^8_\text{comp}$ in PM equilibrium states at $T=0.033$ (circles) and $T=T_\text{eff}=0.38$ (crosses) for interactions $U-dU,U+dU,U$ in orbitals $\alpha=1,2,3$ ($U=8$). The dashed line shows the value in the initial I,AFM,C state. 
}
\label{fig_susc}
\end{center}
\end{figure}

To study the susceptibility for orbital symmetry breaking, we apply slightly different interactions to the three orbitals: $U-dU$, $U$, $U+dU$ for $\alpha=1,2,3$. While the low-temperature PM insulating state has a very large (but non-diverging) susceptibility (see line with circles in Fig.~\ref{fig_susc}) this is not the case at the effective temperature $T_\text{eff}=0.38$ of the photo-doped state (line with crosses). This suggests that the susceptibility for orbital symmetry breaking does not play a dominant role in the stabilization of the hidden state. 

\end{document}